\documentstyle[preprint,revtex]{aps}
\begin{document}
\draft
\preprint{LSUHE No. 185-1995}
\def\overlay#1#2{\setbox0=\hbox{#1}\setbox1=\hbox to \wd0{\hss #2\hss}#1%
\hskip -2\wd0\copy1}
\begin{title}
Dual Abrikosov vortices in U(1) and SU(2) lattice gauge theories
\end{title}
\author{Richard W. Haymaker}
\begin{instit}
Department of Physics and Astronomy, \\
Louisiana State University, Baton Rouge,
Louisiana 70803-4001, USA

\end{instit}
\begin{abstract}

The spatial distribution of fields and currents
in confining theories can give direct evidence of dual
superconductivity.  We review the behavior of vortices in
the lattice Higgs effective theory.
We discuss the techniques for finding these properties
and calculating the superconductivity parameters in lattice
simulations.  We have seen dual Abrikosov vortices directly in pure
U(1) and SU(2) and others have also seen them in SU(3).
We review the duality transformation for U(1) in order to
connect the U(1) results to a dual Higgs theory.
 In the non-Abelian
cases the system appears to be near the borderline between type I and II.
We also discuss the response of the  persistent currents to
external fields.

\end{abstract}

\pacs{\\ \\ Lectures presented at the International School of Physics
``Enrico Fermi" Course CXXX: ``Selected topics in non perturbative QCD",
Varenna, Villa Monastero, 27 June - 7 July 1995.  \\  \\  \\
%PACS number(s): 11.15.Ha
 }

\narrowtext
\section{Introduction}
\label{intro}

The physics of superconductivity has provided the inspiration for many of
the ideas behind the color confinement mechanism in QCD.
The most notable property is
the behavior of a magnetic field imposed normal to the surface of
such a material.  For a type II superconductor,
the field drills holes in the superconductor forming
a lattice of filaments of normal material defining the cores of
Abrikosov\cite{abrikosov} vortices\cite{tinkham}.
The longitudinal magnetic field is a
maximum at the center
and falls off exponentially in the direction transverse to the
core over a distance denoted as the London penetration depth, $\lambda$.
One quantum of magnetic flux is carried by each vortex.  Recent photographs
using a scanning-tunneling-microscope\cite{belllabs} and
electron-holograph-interferometry\cite{hitachi} show this
flux confinement very dramatically.

Dual superconductivity\cite{back1} is the analogous phenomenon in which
persistent currents of
magnetic monopoles form dual Abrikosov vortices which confine
electric flux.
Lattice gauge theory offers the prospect of
exploring dual superconductivity
 in depth as a confining mechanism.
 In four dimensional U(1)
lattice gauge theory there is considerable indirect evidence from
``bulk properties'' of the vacuum such as the monopole
density~\cite{dt}, monopole susceptibility~\cite{cardy}, and static
quark potential~\cite{ws} that Dirac magnetic monopoles are associated
with the phenomenon of confinement.

The existence of a dual Abrikosov vortex
between a static quark-antiquark pair leads to  more direct evidence for
confinement.  By studying the relationship between
the monopole currents and
fields in the neighborhood of static sources we identified the
flux tube with a dual vortex in U(1)\cite{shb,hsbw}, and in
SU(2)\cite{sbh}.  Suzuki and collaborators reproduced the SU(2)
results with better statistics and verified the effect
in SU(3)\cite{mes}.  We have further studied the effect at finite
temperature and have seen the vortex disappear above the deconfining
temperature\cite{ph}.  Cea and Cosmai\cite{cc} have measured the
electric field profiles in a flux tube in U(1) and SU(2) identifying
the shape with a dual vortex.

In ordinary superconductivity, the primary issue is the spontaneous breaking
of the electromagnetic U(1) gauge symmetry (SSB)
signaled by the non-vanishing of
the vacuum expectation value of a charged field.
An immediate consequence
is that the curl of the vector potential is proportional to the curl
of the electric current known as  the London relation\cite{london}.
 The London relation is violated only near the boundaries of
superconducting material
within a distance defined as the Ginzburg-Landau\cite{gl} coherence length,
$\xi$.
Combining the London relation with Maxwell's equations gives
mass to the electromagnetic field.  The Meissner\cite{meissner}
 effect and infinite
conductivity follow.

These relationships
suggest numerous ways to search for signals of dual superconductivity.
Numerical methods may be successful in establishing some of these
connections and not others.  The focus of this work is to pursue
the dual London equation which is central to the phenomenon.  There are
other promising approaches which we will also describe.

The theory of superconductivity entails (i) the identification of
what symmetry is broken and what are the relevant dynamical
coordinates, (ii) the mechanism that leads to the
the instability and hence  SSB, e.g. BCS\cite{bcs} theory, and then (iii) an
effective theory of the currents and fields in the broken phase, e.g.
Ginzburg-Landau\cite{gl} [GL] theory, or equivalently the four dimensional
generalization, the Higgs model\cite{higgs} treated as an effective theory,
which allows the calculation of the spatial consequences of the
broken symmetry\cite{tinkham}.

Our goals are similar to these.
We hope to identify relevant dynamical coordinates and a corresponding
effective theory.   Experience has shown that an analytic approach
is very difficult.
The duality transformation has not been
achieved for the Wilson form of the action which we are using.
But this transformation has
been implemented in closed form in the
closely related Villain\cite{villain} form of the action by
Polyakov\cite{polyakov} and Banks, Myerson and Kogut\cite{bmk} in the
early development of this subject.

Fr\"{o}lich and Marchetti\cite{fm}  have developed these techniques
further.  They have identified the monopole field operator
for the Villain action.  The non-vanishing  vacuum expectation value
is identified as the order parameter for dual superconductivity.
 This has been studied
numerically by Polley and Wiese, and Polikarpov\cite{pw,ppw}.
Ref\cite{bwpp,zpw} are other closely related papers.

Similar analytic methods have not been successful for the
Wilson form for the U(1) action.  However
Del Debbio, Di Giacomo, Paffuti and Pieri\cite{giacomo2}
have identified the monopole field operator applicable
for a more general U(1) action and found the vacuum
expectation value to be a very strong signal for dual superconductivity.

While the origin of confinement in U(1) lattice gauge theory is fairly
clear,  understanding confinement in non-Abelian SU(N)
theories has been more difficult~\cite{tHooft1}.
 One promising approach is to
fix the non-Abelian degrees of freedom in the maximal Abelian
gauge as advocated by Schierholz and Suzuki and others~\cite{klsw,sy},
leaving a residual U(1)$^{N-1}$ gauge freedom,
with ($N$-1) species of U(1) Dirac monopoles.
Suzuki and collaborators\cite{suzuki,suzuki1} have laid much of the
foundation for this lattice approach. The monopoles have been
observed to be abundant in the confined phase
and dilute in the (finite temperature)
unconfined
phase\cite{klsw,sy,suzuki,suzuki1,born,woloshyn,giacomo1,bsw,ss,ipp}.
A number of contributions by
Yee\cite{yee2,yee3,yee1} are described in a conference talk\cite{yee2}.

Del Debbio, Di
Giacomo, Paffuti and Pieri\cite{giacomo2} have applied their U(1) construction
to the Abelian projected SU(2) and again saw a clear signal of dual
superconductivity.  They have considered a number of forms of the
Abelian projection but have not yet implemented this method for
the maximal Abelian gauge.

Dual vortices have been seen for SU(2) as reported in Refs.\cite{sbh,mes,ph}.
All these papers report
the similar conclusion that the dual GL coherence length $\xi_d$ and
the dual London penetration depth $\lambda_d$ are roughly equal. For
a type II dual superconductor
$\kappa_d \equiv \lambda_d/\xi_d > 1/\sqrt 2$,
 and a type I otherwise\cite{tinkham}.  Maeda, Matsubara and
 Suzuki\cite{mms} came to a similar conclusion for SU(3) from an
 entirely independent argument.  For superconducting materials
the difference between type I and type II is dramatic when a magnetic
field is imposed.  The penetration of magnetic flux in a type I material
leads to very complex patterns of normal and superconducting domains
called an 'intermediate state structure'.  Numerous photographs of this
phenonenon are shown in a monograph by R. P. Huebener\cite{huebener}.

Since U(1) lattice gauge theory does not have an non-trivial continuum
limit, one must use the results with caution in describing confinement
in QCD.  However it is more than just an independent model of confinement.
Yee\cite{yee1} has found using the demon method
that for small $\beta$ the
 Abelian projected action for SU(2) is essentially the U(1) action with
$\beta_{U(1)} = \beta_{SU(2)}/2$.  Further this must
and  indeed Yee finds  it does break down in an interesting way as $\beta$
approaches the scaling region. See also Suzuki\cite{suzuki}
 for a different approach.

In a long series of papers and a review article,
Baker, Ball, Zachariasen\cite{bbz}
and co-workers have postulated a dual form of
QCD in the continuum.  Their formulation describes a non-Abelian dual
superconductor and hence it confines color. They have calculated
flux tubes, static quark potentials,  temperature dependent effects
and many other quantities in the tree approximation.  Although
their formulation does not focus on monopoles, there is a large
potential overlap between their work and the lattice description
developed here.

In Section\ref{higgs} we review the Abrikosov vortex in the
Higgs theory\cite{higgs} which is the basis of the many interrelated
signals of superconductivity.   We will emphasize the lattice form of
the Higgs model.  The singularities giving vortices arise in an
interesting way naturally on the lattice.

In Section\ref{u1} we report the results of numerical studies of U(1) for
correlations between a Wilson loop and a plaquette giving a very clear
signal for the London penetration depth.  These numerical results
suggest that a dual form of the model should exhibit this phenomenon
in a simple way.  In Section\ref{duality} we show that the simpler
Villain\cite{villain}
form of the action does allow an analytic treatment of the question and
we review the results of Fr\"{o}lich and Marchetti\cite{fm},
Polley and Wiese\cite{pw},
and Polikarpov\cite{ppw}.

Results for SU(2) and SU(3) are given in Section\ref{su2}.  The analysis
involves an approximate treatment of the GL theory in order account
for the behavior of the fields in the dual vortex.

Fields and currents exist only near the boundaries of dual superconductors
over a distance scale set by the London penetration depth $\lambda_d$.
It is these
spatially transient effects that provide spatial structures where
the local properties of dual superconductivity can be studied and that
is the focus of this work.  One way to study these phenomena is to
impose an external field.  In Section\ref{efield}  we introduce
an external field and discuss some of the prospects and problems.

The lattice approach provides a very concrete way to visualize physical
quantities as living on links, plaquettes,  dual plaquettes etc.  The
use of the abstract language of differential forms is gaining use
in the lattice literature.  This formulation focus on k
dimensional volumes and boundaries of the volumes in a systematic
way.  The technology in some ways removes one from the
nitty gritty of the lattice but with some familiarity  it
actually enhances ones ability to see general relationships.  Polley
Weise\cite{pw} give a brief introduction to the basics of this
formulation.   In the appendix
we go one step further and give a concrete translation
of differential forms into lattice difference operators.

\bigskip
\section{Abrikosov vortices}
\label{higgs}

The Ginzburg-Landau effective theory of superconductivity is a time
independent description based on a complex order parameter, $\psi(\vec{x})$,
where $|\psi(\vec{x})|^2$ is the number density of superconducting
charge carriers.  In normal material $|\psi(\vec{x})| = 0$.
We will focus on the
special circumstance in which $|\psi(\vec{x})|$
is constant in the superconducting
phase and discontinuous at the surface. This is the extreme type II limit,
also called the London limit.

The Higgs\cite{higgs} model is the four dimensional generalization of
GL theory with the Higgs field $\phi(x)$ corresponding to the GL order
parameter $\psi(\vec{x})$.  The London limit is obtained by constraining the
Higgs field:
\begin{equation}
V_{Higgs}(|\phi(x)|^2) \Rightarrow |\phi(x)|^2 = v^2.
\label{e5.1}
\end{equation}
In other words the London theory is equivalent to
the effective non-linear Higgs theory.

\bigskip
\subsection{London theory}
\label{london}

A concise statement of the London\cite{london} theory is contained
in the relation
\begin{equation}
\vec{A} + \lambda^2\vec{J}= 0; \;\;\;\;
(\vec{\nabla} \cdot \vec{A} = 0).
\label{121}
\end{equation}
If the charge density is zero, then in this gauge the electric field
is given by $-\dot{A}$ and therefore
$\vec{E} = \lambda^2 \dot{\vec{J}}$.
This describes a perfect conductor and is just Newton's law for
free carriers, $e \vec{E} = m \dot{\vec{v}}$.  By taking the curl
of Eqn.(\ref{121}) we obtain the condition for a perfect diamagnet.
\begin{equation}
\vec{\nabla} \times \vec{J}  = - \frac{1}{\lambda^2} \vec{B}.
\label{140}
\end{equation}
This relation together with Ampere's law
$\vec{\nabla} \times \vec{B} = \vec{J}$ gives
$\nabla^2 \vec{B} = \vec{B}/\lambda^2$ which implies that the magnetic field
falls off in the interior of the superconductor with a skin depth $\lambda$.

Finally the fluxoid is given by the integral
\begin{equation}
\int_{S} (\vec{B}+ \lambda^2\vec{\nabla} \times \vec{J})\cdot \hat{n} da =
\int_{C} (\vec{A}+ \lambda^2\vec{J})\cdot d\vec{l} =
N \frac{2 \pi}{e} =N e_m = \Phi_{m}.
\label{150}
\end{equation}
If the curve $C$ is in a simply connected region of a superconductor, then
$N = 0$.  However if the curve encircles a hole in the material then
N need not be zero but must be an integer due to the quantization of flux.
 In an extreme type II
Abrikosov vortex, a very
small core is comprised of normal material.  A single unit of magnetic
flux of radius $\sim \lambda$ passes through the vortex.  The fluxoid
density $\vec{B} + \lambda^2\vec{\nabla} \times \vec{J}$ is zero
everywhere except in the region of the normal material.  In the limit
in which the core is a delta function we obtain:
\begin{equation}
\vec{B} + \lambda^2\vec{\nabla} \times \vec{J}=
\Phi_{m} \delta^2(x_{\perp})\hat{n}_z.
\label{160}
\end{equation}
Further if we use Ampere's law we can get an analytic expression for the
$B_z$ profile of a vortex.
\begin{equation}
B_z(r_{\perp}) - \lambda^2\nabla_{\perp}^2 B_z(r_{\perp})=
\Phi_{m} \delta^2(x_{\perp});
\;\;\;\; B_z = \frac{\Phi_{m}}{2 \pi \lambda^2} K_{0}(r_{\perp}/\lambda).
\label{170}
\end{equation}

\bigskip
\subsection{Lattice Higgs in the London limit}
\label{latticehiggs}

I would like to review the salient features of the lattice
Higgs theory in the tree approximation.
It is interesting to see how the
the London relation arises as a direct consequence of the
spontaneous breaking of the U(1) gauge theory.  Also we compare
the simulations
of pure gauge theories with this model interpreted as a dual theory.

Consider the lattice action:
\begin{equation}
S = \beta \sum_{x,\mu > \nu} (1 - cos\;\theta_{\mu\nu}(x)) -
    \kappa \sum_{x,\mu}(\phi^{*}(x)e^{i\varphi_{\mu}(x)}
                        \phi(x+\epsilon^{(\mu)}) + H.c.) +
    \sum_x V_{Higgs}(|\phi(x)|^2 ).
\label{e1}
\end{equation}

We need a background field in order to form an Abrikosov vortex.
  To keep as close as possible to our simulations we
generate the field from a monopole loop, i.e. the dual analog
of the Wilson loop.  Hence we will digress to define the monopoles on
the lattice using the DeGrand-Toussaint\cite{dt} construction.

Consider the unit 3-volume on the lattice at fixed $x_4$ as shown in
Fig.\ref{fdt}.   The link angles are compact,
$-\pi < \varphi_{\mu} \leq \pi$.
The plaquette angle  is also compact,
$-4\pi < \theta_{\mu\nu} \leq 4\pi$ and defined
\begin{equation}
   ea^2 F_{\mu\nu}(x)= \theta_{\mu\nu}(x) =
              \Delta^{+}_{\mu} \varphi_{\nu}(x) -
                        \Delta^{+}_{\nu} \varphi_{\mu}(x),
\end{equation}
where $a$ is the lattice spacing and the
difference operators are defined
in the appendix.
This  measures the electromagnetic flux through the face.
Consider a configuration in which all link
angles on this cube are small compared to $\pi$.
 Gauss' theorem
applied to this cube then clearly gives zero total flux.
Because of the $2\pi$ periodicity of the action we
decompose the plaquette angle into two parts
\begin{equation}
   \theta_{\mu\nu}(x) = \bar{\theta}_{\mu\nu}(x) + 2 \pi n_{\mu\nu}(x).
 \label{thetabar}
\end{equation}
where $-\pi < \bar{\theta}_{\mu\nu} \leq \pi$.
If
the four angles shown in Fig.\ref{fdt} are adjusted so that e.g.
$\theta_{\mu\nu}>\pi$ then
there is a discontinuous change in $\bar{\theta}_{\mu\nu}$  by $-2 \pi$
and a compensating change in $n_{\mu\nu}$.
  We can
clearly choose the configuration that leaves the plaquette angles on
all the other faces safely away from a discontinuity.  We then define
a Dirac string $n_{\mu\nu}$ passing through this face (or better a
Dirac sheet since the lattice is 4D).
 $\bar{\theta}_{\mu\nu}$ measures the electromagnetic flux
through the face.
The important points are the following:
\begin{itemize}
\item The volume shown contains a magnetic monopole as indicated
  by a net flux of $2\pi/e = e_m$.
\item Note that if one does a gauge transformation e.g. at the point
     $x + \epsilon^{(\mu)} +  \epsilon^{(\nu)}$ then if one of the affected
     links picks up a discontinuity of $2 \pi$, the two contiguous
     plaquettes will be affected but the flux out of the volume is
     unchanged.  The monopole is a gauge invariant construction.
\item This example created a monopole-antimonopole pair in the two
      space cubes separated by a lattice spacing in the $x_3$ direction.
      The same holds in the $x_4$ direction (not shown in fig.\ref{fdt}).
\item  In the dual description, cubes in the original lattice
       are represented  by links on the dual lattice orthogonal
       to the 3-volume.
\item  Hence for $n_{\mu\nu} = 1$, (a Dirac sheet threading the plaquette)
       a closed $1 \times 1$ monopole loop
       has been created on the dual lattice in the $x_3,  x_4$ plane
       forming the boundary of a Dirac sheet.
\item  The monopole loops are closed, hence the current is conserved.
\end{itemize}
This construction gives the following definition of the magnetic monopole
current.
\begin{equation}
\frac{a^3}{e_m}J^m_{\mu}(x)= \epsilon_{\mu \nu \sigma \tau}\Delta^+_{\nu}
\bar{\theta}_{\sigma \tau}(x)
 \label{magcur}
\end{equation}
Which lives on the dual lattice.   It satisfies the
 conservation law $\Delta^+_{\mu}J^m_{\mu}(x) = 0$.

 We obtain a classical solution by minimizing
the action.  Our initial configuration contains a single
closed magnetic monopole loop in the $x_3, x_4$ plane analogous to
the Wilson loop projector in pure gauge simulations.
This can be easily accomplished by choosing a configuration with
a Dirac sheet threading  all 1-2 plaquettes in a the 3-4 plane
that spans the desired monopole loop.
Our algorithm then rejects all updated links that change
the initial monopole configurations.

We accept configurations with Dirac sheets with no accompanying
monopoles.  A lump of Dirac sheets is rather easily created
in a single local update.
 The update of a single link can
produce six Dirac sheets in the plaquettes contiguous to this link.
Each of these produces a monopole loop on the dual lattice.  One can
see  as shown in Fig.\ref{fdual}
 that the six  dual monopole loops form the
faces of a cube on the dual lattice in such a way that all
the monopole currents on the
edges of the cube cancel.

We use the method of simulated annealing,
slowly increasing $\beta$, holding  $ \lambda^2/a^2 = \beta/\kappa
= 1/e^2 \kappa$
constant, where $\lambda/a$ is the
London penetration
depth in lattice units.

The electric current is given by
\begin{equation}
 \frac{a^3}{e \kappa} J^{e}_{\mu}(x) =
Im ( \phi^{*}(x)e^{i \varphi_{\mu}(x)}
                                        \phi(x+\epsilon^{(\mu)})).
\label{e2}
\end{equation}
Writing the Higgs field $\phi(x) = \rho(x) e^{i \omega(x)}$,
the curl of the current is hence given by
\begin{eqnarray}
\lefteqn{ \frac{a^4}{e \kappa}\frac{(\Delta^+_{\mu} J^{e}_{\nu}(x)
                    - \Delta^+_{\nu} J^{e}_{\mu}(x))}{a}  =}\nonumber\\
             &&       \rho(x) \rho(x+\epsilon^{(\mu)})
                     \sin[ - \omega(x)+ \varphi_{\mu}(x) +
                             \omega(x+\epsilon^{(\mu)})]+ \nonumber \\
               &&   \rho(x+\epsilon^{(\mu)})
                     \rho(x+\epsilon^{(\mu)}+\epsilon^{(\nu)})
                     \sin[ - \omega(x+\epsilon^{(\mu)})+
                              \varphi_{\nu}(x+\epsilon^{(\mu)}) +
                     \omega(x+\epsilon^{(\mu)}+\epsilon^{(\nu)})]- \nonumber \\
               &&   \rho(x+\epsilon^{(\nu)})
                     \rho(x+\epsilon^{(\nu)}+\epsilon^{(\mu)})
                     \sin[ - \omega(x+\epsilon^{(\nu)})+
                              \varphi_{\mu}(x+\epsilon^{(\nu)}) +
                    \omega(x+\epsilon^{(\mu)}+\epsilon^{(\nu)})]- \nonumber \\
               &&   \rho(x) \rho(x+\epsilon^{(\nu)})
                     \sin[ - \omega(x)+ \varphi_{\nu}(x) +
                             \omega(x+\epsilon^{(\nu)})],
\label{e3}
\end{eqnarray}
Compare this with the electromagnetic field tensor
\begin{equation}
ea^2 F_{\mu \nu}  = \sin[  \varphi_{\mu}(x)
                         + \varphi_{\nu}(x+\epsilon^{(\mu)})
                         - \varphi_{\mu}(x+\epsilon^{(\nu)})
                         - \varphi_{\nu}(x)].
\label{e4}
\end{equation}
If the U(1) gauge symmetry is spontaneously broken, these two
quantities are equal which is the London relation.   To be more precise:
if (i) $\rho$ is nonvanishing and independent of position (absorb the
normalization into $\kappa$) and (ii)
\begin{equation}
 \sin[\theta + 2N\pi] \approx \theta,
\label{e4.1}
\end{equation}
 then\footnote{An apparent sign difference with Eqn.(\ref{121}) is a
notational one, see the appendix \ref{maxwell}.}
\begin{equation}
{\cal F}_{\mu \nu} \equiv  F_{\mu \nu}
   - \frac{a^2}{e^2 \kappa}\frac{(  \Delta^+_{\mu} J^{e}_{\nu}(x)
                            - \Delta^+_{\nu} J^{e}_{\mu}(x))}{a}
     = \frac{2\pi N}{e} \frac{1}{a^2}
    =  N e_m \frac{1}{a^2},
\label{e5}
\end{equation}
${\cal F}_{\mu \nu} a^2$ summed over the transverse plane
is called the fluxoid and hence ${\cal F}_{\mu \nu}$
is the fluxoid density.  The magnetic charge $e_{m}$ is the
Dirac monopole charge.  Condition (i) is satisfied by freezing out the
modulus of the Higgs field with the constraint Eqn.(\ref{e5.1}).
and (ii) is satisfied if the variation of the field modulo $2 \pi N$
is much
smaller than a lattice spacing.  (Note that this singularity is not an
isolated Dirac sheet. That would act differently, adding  $2 \pi n$ to both
terms in Eqn.(\ref{e5}) and giving a zero contribution on the right.)

Eqn.(\ref{e5}), together with Maxwell's equations, gives the lattice
version of the Meissner
effect: $ \vec{B} - \lambda^2 \Delta \vec{B} = 0$, where
$ \lambda^2 = m_{\gamma}^{-2}$;
infinite conductivity
$\vec{E} = \lambda^2 \Delta^+_{4}\vec{J}$ (assuming
$\rho = 0$); and an Abrikosov vortex
\begin{equation}
B_z - \lambda^2(\vec{\Delta}^+\times \vec{J})_z =
 N \frac{e_m}{a^2}  \delta^2_{x_{\perp},\; 0}.
\label{e5.2}
\end{equation}

Fig.\ref{f1} shows the profile of the R.H.S.
 of eqn.(\ref{e3}) and eqn.(\ref{e4}) in the
directions perpendicular to the $5 \times 5$ magnetic monopole loop on a
$12^4$ lattice, with $\beta/\kappa = 1$. We used
the constrained form of the Higgs potential, eqn.(\ref{e5.1}),
which corresponds
to an extreme type II superconductor.
The graphs show the expected behavior: i.e. the equality of the two quantities
everywhere except at $r=0$ where they should differ by $2\pi$. There are
significant violations only at $r=0$ and smaller violations at $r=1$ due
 to the breakdown of Eqn. (\ref{e4.1}).  We anticipate that the
violations will decrease with increasing loop size.
This solution is a possible way
to treat the complications that are ignored in the continuum London
theory described in Sec.\ref{london}

Note that there is a sign change in the $curl J^e$ profile.  Eqn.(\ref{150})
shows that there must be a sign change since it must integrate to zero
over the plane.  The vortices have an exponential profile.  All the current
circulates about the core in the same sense.  However since the profile
falls faster than $1/r$, the line integral around a small patch at the
 origin will have
the opposite sign than the line integral around a small patch elsewhere.

It is interesting to note that the interior surface
spanned by the monopole loop is in the normal phase since the London relation
is violated there.
All other regions are superconducting.
This translates in the following
sections in which the Wilson loop projects out the normal phase  in the
plane spanned by the loop and the
remaining regions are a dual superconductor.

We chose the Higgs field $|\phi(x)| = 1$ for this discussion. If instead
the field was subjected to a SSB potential
\begin{equation}
V_{Higgs}(|\phi(x)|^2) = \lambda (|\phi(x)|^2 -v^2)^2 ,
\label{mexican}
\end{equation}
there are
important differences in the solution. In the region near a superconducting
normal boundary, one expects a soliton like deformation of the Higgs
field, constrained to vanish in the normal region and favoring a
non-vanishing expectation value deep inside
the superconducting side. This
transition region defines the Ginzburg-Landau coherence length $\xi$,
 which is the
distance over which the order parameter rises to its asymptotic value.
For distances into the superconducting material deeper than the
coherence length, one expects the London relation to be satisfied.

\section{Dual superconductivity in pure U(1) gauge theory}
\label{u1}

We now turn to the dynamical simulations in pure U(1) gauge theory given by
\begin{equation}
S = \beta \sum_{x,\mu > \nu} (1 - cos\;\theta_{\mu\nu}(x)).
\label{eu1}
\end{equation}
We have given evidence
ref.\cite{shb} for the dual fluxoid density relation signaling
a dual Abrikosov vortex:
\begin{equation}
{\cal ^*F}_{\mu \nu} \equiv \; ^*F_{\mu \nu}
   - \lambda_{d}^2\frac{(  \Delta^-_{\mu} J^m_{\nu}(x)
                            - \Delta^-_{\nu} J^m_{\mu}(x))}{a}
     =  \frac{e}{a^2} \delta_{x_{\mu},0} \; \delta_{x_{\nu},0}.
\label{e6}
\end{equation}
where $ ^*F_{\mu \nu} = -\frac{1}{2}\epsilon_{\mu \nu \sigma \tau}
F_{\sigma \tau}$.
The magnetic monopole current is defined in Eqn.(\ref{magcur}).

The correlation of $\vec{\Delta} \times \vec{J}^m$
with a Wilson loop gives this signal for the solenoidal behavior
of the currents surrounding the electric flux between oppositely charged
particles.
The correlators for the electric field  $\vec{E}$ and
$\vec{\Delta^-} \times \vec{J}^m$ are
\begin{eqnarray}
e a^2 E_z &=&
    -\frac{<e^{i \theta_{W}}  e^{i \theta_{P}}>}
         {<e^{i \theta_{W}}>} =
   \frac{<\sin{\theta_{W}} \sin{\theta_{P}}>}
         {<\cos{\theta_{W}}>}  \nonumber\\
\frac{a^4}{e_m}(\vec{\Delta^-} \times \vec{J}^m)_z &=&
   -\frac{<e^{i \theta_{W}} \;\;
 i(\vec{\Delta^-} \times \vec{J}_{lat}^m)_z>}
{<e^{i \theta_{W}}>} =
     \frac{<\sin{\theta_{W}}\;\;
(\vec{\Delta^-} \times \vec{J}_{lat}^m)_z>}
{ <cos{\theta_{W}}>}
\label{cor}
\end{eqnarray}
The notation $(\cdots)_{lat}$ denotes an integer valued lattice quantity.

The identification of $sin{\theta_P}$ with the electric field is
arbitrary, any quantity with the correct naive continuum limit will
do.   However Zach, Faber, Kainz and Skala made the interesting
observation that this choice in conjuction with the Wilson
action satisfies Gauss' law\cite{zfks}.

For $x_{\mu}$ out of the plane of the Wilson loop, the RHS of
Eqn.(\ref{e6}) vanishes and the London relation is satisfied.  Hence
we interpret the Wilson loop as a projector onto the normal phase
in the plane bounded by the loop and a dual superconductor elsewhere.

Fig. \ref{curl} shows the lattice  operators for the electric
field and the curl of the magnetic monopole current.
The longitudinal electric
field, (a), is given by a $z,t$ plaquette which is
depicted by a bold line for fixed time.
The curl of the magnetic monopole current, (b), is built from four
3-volumes which appear as squares since the time dimension is not shown.
Passing through the center of each square is the link dual to the 3-volume.
One takes the monopole number $n$ in each 3-volume and associates the value
$n e_{m}$ with the corresponding dual link.
The `line integral' around the dual plaquette
completes the picture. Notice
from this construction that $\vec{E}$ and $\vec{\Delta^-}\times \vec{J}^m$ take
values at the same location
within the unit cell of the lattice, both are indicated by the
bold face line in the $z$ direction.
Both operators live on the same $z,t$ plaquette.  However
$\vec{\Delta^-}\times \vec{J}^m$ is a hefty operator involving many
links and as a consequence gives lattice artifacts for correlations
at modest separations.

Fig.\ref{f3} shows the profile for the quantities making up the London
relation on the same plot as for the effective lattice Higgs case,
Fig.\ref{f1}.  We did 800 measurements on a $12^4$ lattice,
 $\beta = 0.95$, and a
$3\times3$ Wilson loop was used to project onto the $q,\bar{q}$ sector
using methods described in Ref.\cite{shb}.

Fig.\ref{f3a} shows the behavior of the dual fluxoid density.
We did a $\chi^2$
fit of the London relation, eqn.(\ref{e6}),
 using the $R \neq 0$ points and determined
$\lambda_d = 0.49(3)$.  The fit strongly favors the small
$R$ points. That determined the $R=0$ value of the dual fluxoid in
eqn.(\ref{e6}), giving a prediction of
$ e = 1.07(7)$, compared to the expected
value $1/\sqrt{\beta} = 1.03$.  All these features correspond directly
to the classical solution of the lattice Higgs model reinterpreted as a
dual theory.

Our goal here is to learn how to measure
 parameters of the dual
superconducting medium.  The Wilson loop projects out a normal region
giving a normal-dual superconducting boundary and a chance to discover
a London relation in the spatially transient region on the dual
superconducting side.  The U(1) model has a rather small correlation
length for $\beta \approx 1$  and it is difficult
to probe large distances.  One small improvement is to replace the
Wilson loop projector by a plaquette and study the tails of the
correlators, Eqn(\ref{cor}).   We present some preliminary
observations but this small scale simulation is inadequate to
tie down a consistent determination of $\lambda_d$.

Fig.\ref{f3b} shows such a plot.  This represents 2000 measurements for
$\beta = 1.0$ and for a plaquette projector.
An approximately exponential fall off is expected.
The curve is only to
guide the eye, it is an eyeball fit to the tail of the electric field:
\begin{equation}
E(R) =  E_0\frac{e^{-R/R_0}}{R^{3/2}}
\end{equation}
where
$E_0 = 0.2$ and  $R_0 = 0.7$.
We use the 4-dimensional behavior for the tail for these 'point like'
operators.

One can read off the point by point determination of the dual London
penetration depth from Fig.\ref{f3b}:
\begin{equation}
 \lambda = \sqrt{\frac{E}{2 \pi [curl J^m ]}}
\end{equation}
On axis, at R = 0 we have instead
\begin{equation}
 \lambda = \sqrt{\frac{E}{2 \pi [curl J^m  E/(E-\Phi_m)]}}
\end{equation}
and the quantity in square brackets is plotted instead of
$[curl J^m (\Box)]$ in order that the  $R = 0$ determination of $\lambda_d$
can be read from the graph the same as the $R \neq 0$ cases.

If we look along a lattice axis transverse to the
projector plaquette , i.e. integer R, then
$ E/curl J^m \approx 1.0$.
giving $\lambda_d \approx 0.4$. compared
to the eyeball fit of $0.7$.  Off the lattice
axis, $R$ noninteger, $ E/curl J^m \approx 2.0$ giving
$\lambda_d \approx 0.56$.  We expect the point by point
determinations of $\lambda_d$
should be consistent with each other and with the exponential tail
determination only at large distances.  This small scale simulation
is inadequate to make a thorough check of this.

A run of 2420 sweeps at $\beta = 0.95$ gave similar results.  An
eyeball fit to the $E$ data  gave
 $ E_0 = 0.4$ and $R_0 = 0.5$ which is consistent
with the Wilson loop determination above.  This agrees with
the expected result that $\lambda_d$ decreases with decreasing $\beta$.

\section{Duality Transformations in U(1)}
\label{duality}

We arrived at the London relation in the U(1) theory purely
numerically in Section \ref{u1}.  This begs the question of whether
the dual formulation of U(1) lattice gauge theory is equivalent to
a lattice Higgs theory with a spontaneously broken gauge symmetry.
Fr\"{o}lich and Marchetti\cite{fm} have found this to be the case which
we review briefly in this section.

The duality transformation has not been
performed for the Wilson form of the action which we are using.
But this transformation has
been implemented in closed form in the
closely related Villain\cite{villain} form of the action by
Fr\"{o}lich and Marchetti\cite{fm} and developed further by
Polley and Wiese, and  Polikarpov
and others\cite{pw,ppw,bwpp,zpw}.
We draw on these later papers to gain more insight into the
dual London relations.

Recent literature on this subject have found it very convenient to use
the differential form notation adapted to the lattice.
Polley and Wiese\cite[Sec. 2.1]{pw} have given a brief
introduction to the subject.  The advantage
of this formulation is that it forces one to focus on the k-volume
cell on the lattice for objects with k indices.  Further it
dispenses with the indices and reduces the typical
manipulations to some very
simple  general operator relations.
 As a supplement to Polley and Wiese's
introduction,  we give the reader a concrete realization of the
differential form algebra in the appendix \ref{diff}
in order to make the transition to this notation a little easier.

Consider the Wilson action
\begin{equation}
Z = \prod \int^{\pi}_{-\pi} d \varphi \;\; \exp\{- \beta \sum_{d\varphi}
(1 - \cos d\varphi)\},
\end{equation}
where the link angle $\varphi$ is a 1-form and the plaquette
$d \varphi$ is a 2-form constructed with the exterior differential, d.
We can expand each plaquette in a fourier series
\begin{eqnarray}
\exp\{-\beta(1 - \cos z)\} &=& \sum_k e^{ik z} e^{-\beta} I_k(\beta),\\
       &\approx& \sum_k e^{ik z}  e^{-k^2/(2\beta)},
\end{eqnarray}
where we have used the asymptotic form of the Bessel function
 $I_k(\beta)$ for large $\beta$ and further for large k.
This gives the Villain\cite{villain} form of the action
\begin{equation}
Z = \prod \int^{\pi}_{-\pi} d \varphi \;\;
     \prod \sum_k \;\; \exp\{i(k,d\varphi) -\frac{1}{2 \beta}
\parallel k \parallel^2\},
\end{equation}
where  the inner product, $(\Phi,\Psi)$, is defined in the
appendix\ref{diff} and
$\parallel \Phi \parallel^2 \equiv (\Phi,\Phi)$.
We now use the property of the inner product, $(k, d \varphi)
= (\delta k, \varphi)$.  We can now integrate over the links $\varphi$
and we obtain
\begin{equation}
Z =      \prod \sum_{k}
\;\; \exp\{-\frac{1}{2 \beta}
\parallel k \parallel^2\}  \biggr\vert_{\delta k = 0}.
\end{equation}
The variable $k$ is an integer valued 2-form  analogous to
the field tensor $F_{\mu \nu}$  and the constraint $\delta k = 0$
is analogous to $\partial_{\mu}F_{\mu\nu} = J^{e}_{\nu} = 0$. Next, using
the general operator relation $\delta^2 = 0$ we
solve the constraint by introducing
\begin{equation}
k = \delta p,
\label{kp}
\end{equation}
\begin{equation}
Z =      \prod \sum_{p}
\;\; \exp\{-\frac{1}{2 \beta}
\parallel \delta p \parallel^2\} \;\;\;\; = \;\;\;\;
  \prod \sum_{^*p}
\;\; \exp\{-\frac{1}{2 \beta}
\parallel d ^*p \parallel^2\}.
\label{deltap}
\end{equation}
The variable $p$ is a  3-form analogous to a dual potential written
as an antisymmetric three index object, $B_{\alpha\beta\gamma}$
and Eqn.(\ref{kp}) is analogous to $F_{\mu \nu} = \partial_{\alpha}
B_{\alpha \mu \nu}$. If we switch to the dual objects this last result looks
much more familiar: $*p$ is a 1-form,
$ ^*k = d ^*p  \;\;\Leftrightarrow \;\;^*F_{\mu\nu} = \partial_{\mu}^*B_{\nu} -
\partial_{\nu}^*B_{\mu}$.

In order to verify this interpretation we should introduce external
sources and observe how they couple to the fields.  Equation
(\ref{deltap}) naively looks like
a free field theory, except that the variables are integer valued and this
indeed makes it an interacting theory.  One can trace the integer value to
the fact that the link angles $\varphi$ are compact and hence their
Fourier transform is a discrete variable.  But this compactness is also
responsible for the existence of monopoles.

Our immediate concern in this paper is to show that this interacting
theory given by Eqn.(\ref{deltap}) is in fact a dual lattice
Higgs model\cite{fm}.  It is interesting how this
comes about and gives us some insight into the
significance of an integer valued dual
gauge field.

Following references\cite{fm,pw}
we introduce a non-compact 1-form gauge field $^*A$ on the
dual links.  We choose the Higgs field to be compact
0-form $^*\Psi = \exp \{i ^*\chi\}$ on the dual sites constrained
on the unit circle.  Finally we introduce an integer
valued 0-form $^*\ell$ on the dual sites
\begin{equation}
Z = \prod \int^{\infty}_{-\infty} d^*A\;\;
\prod \int^{\pi}_{-\pi} d^*\chi \;\;
\prod \sum_{^*\ell} \exp\{ -\frac{1}{2 e_m^2}\parallel d ^*A \parallel^2
- \frac{\kappa_d}{2} \parallel d ^*\chi + 2 \pi ^*\ell - ^*A\parallel^2\}.
\end{equation}
This differs from the lattice Higgs action, Eqn.(\ref{e1},\ref{e5.1}), in two
respects: (i) The gauge field is non-compact, and
(ii) the Villain\cite{villain} form is employed.  The variables
$^*\ell$ are introduced in order to make the action periodic in $^*\chi$.

To connect this theory to Eqn.(\ref{deltap}) we first choose a gauge
for which $^*\chi = 0$.  Then note that for large $\kappa_d$
 the integral over $^*A$ is peaked at the values $^*A = 2 \pi ^*\ell$.
In the limit $\kappa_d \rightarrow \infty$ this becomes
\begin{equation}
Z =        \prod \sum_{^*\ell}
\;\; \exp\{-\frac{(2\pi)^2}{2 e_m^2}
\parallel d ^*\ell\parallel^2\}.
\label{deltaell}
\end{equation}
Identifying $^*\ell$ with $^*p$ this is the
same as Eqn.(\ref{deltap}) with $1/\beta \equiv e^2 =
(2\pi)^2/e^2_m$.

Following  Sec.\ref{u1}  the dual
London penetration depth in this theory
is given by $\lambda^2_d/a^2 = 1/(m_{\gamma} a)^2 = 1/(e^2_m \kappa_d)$.
Therefore in the weak magnetic coupling limit, $e_m \rightarrow 0$,
the U(1) gauge theory with the Villain form of the action becomes a
dual Higgs theory with an infinite photon mass, i.e. zero London
penetration depth, $\lambda_d$.
We find in our simulations that $\lambda_d$ decreases
for decreasing $\beta$.  But we have only measured it for a very
limited range of $0.9 < \beta  < 1.0 $.

The dual Higgs field, $^*\Psi$ has been connected to the monopole
field operator\cite{fm,pw}.  Polikarpov, Polley and Wiese\cite{ppw}
have
calculated the constraint effective potential for this operator
and showed that there is a global symmetry breaking in the dual
superconducting phase.

\section{Generalization to pure SU(2)and SU(3) gauge theories}
\label{su2}

In ref.\cite{sbh} we applied these same techniques and showed that
dual Abrikosov vortices also occur in SU(2) pure gauge theory in the
maximal Abelian gauge.  Matsubara et.al.\cite{mes}
confirmed these results with
better statistics and
generalized them to SU(3).  The SU(2) link matrices are $U_{\mu}(x)$ and
the action is
\begin{equation}
 S = \beta \sum_{x,\mu > \nu}(1- \frac{1}{2}Tr U_{\mu \nu}(x)).
\label{e7}
\end{equation}
The maximal Abelian gauge is defined by maximizing the quantity
\begin{equation}
 R  =  \sum_{x,\mu}Tr[\sigma_3 U_{\mu}(x)\sigma_3 U_{\mu}^{\dagger}(x)].
\label{e8}
\end{equation}
The Abelian link angle is then taken as the phase of $[U_{\mu}(x)]_{11}$ and
the calculation can proceed with little change\cite{klsw,sy}.

Recently we have done the calculation at finite temperature in order to
check this picture on each side of the deconfining
phase transition\cite{ph} which we report here.
The results are shown in the confining phase, fig.\ref{f4},
$\beta = 2.28$  and  the deconfining phase, fig.\ref{f5},
$\beta = 2.40$ on a lattice $4 \times
17^2 \times 19$,  with 800 measurements for each case. Gauge fixing
required about 600 sweeps for each configuration.

Fig.\ref{f4} shows an important difference from the U(1) case.  Whereas
in the U(1) case a linear combination of $E$ and $curl J^m$ for $r \neq 0$
can vanish giving the London relation, it is clearly not possible here. The
behavior of $-curl J^m$ does not match that of $E$.  We interpret this
discrepancy as a signal of a non-zero Ginzburg-Landau coherence
length, $\xi_d$.
Unlike the U(1) case, the
extreme type II limit, in which the superconducting order parameter turns
on at the surface, the evidence here is that the order parameter turns on
over a distance $\xi_d$.
The value of the coherence length is approximately
the radius where the London relation is restored.

The analysis in this more general situation precludes a point by point
comparison of the $E$ and $curl J^m$ data.  We adapt the analysis from
Tinkham\cite{tinkham}.

 In terms of a Higgs model, the Higgs field is not constrained
to a certain vacuum expectation value but rather is subjected to
a SSB potential Eqn.(\ref{mexican}).  The equivalent GL theory provides
a way to calculate the deformation of the Higgs field $\phi(x)$
between the normal
core of a vortex and the asymptotic value at large distance from the
vortex, $\phi_{\infty}$.
We write $\phi(x) = \phi_{\infty} f(r) \exp{i \alpha(x)}$, where $f(R)$
rises from $0 \rightarrow 1$ over a distance  $\xi_d$.  Introducing
a dual vector potential, $\vec{A_d}$, the dual London equation
 is replaced by the dual GL equations for the azimuthal component of the
monopole current.
\begin{eqnarray}
\vec{E} &=& \vec{\nabla} \times \vec{A_d}, \nonumber \\
\vec{J}^m &=& \frac{f^2(r)}{\lambda_d^2}
(\vec{A_d} - \frac{\Phi_e}{2 \pi} \vec{\nabla}\alpha),
\label{ggll}
\end{eqnarray}
and $f$ satisfies
\begin{equation}
-\xi_d^2 \nabla^2 f + \xi^2
(  \vec{\nabla}\alpha  - \frac{2 \pi}{\Phi_e}\vec{A_d})^2 f
-f + f^3 = 0.
\end{equation}
The generalized fluxoid relation becomes
\begin{equation}
\vec{E}(r) - \lambda_d^2 \vec{\nabla} \times \biggr(
\frac{\vec{J}^m(r)}{f(r)^2}\biggr)
= N\Phi_e \delta(x_{\perp}).
\label{144}
\end{equation}
An approximate solution of Eqn.(\ref{ggll}) is
\begin{equation}
f(r) = tanh(\nu r/\xi_d),
\end{equation}
with $\nu \approx 1$.

An interpolation of the data is required  to fit Eqn.(\ref{144}).
An example of the fit for $\beta = 2.4$ in Ref.\cite{sbh} is
$\lambda_d/a = 1.05(12)$ and $\xi_d = 1.35(11)$.

The behavior of all the SU(2) and SU(3) examples in the confined phase
is similar to that shown in
fig.\ref{f4}.  The interesting conclusion is that
\begin{equation}
\kappa_d \equiv \lambda_d/\xi_d \approx 1.
\end{equation}
(no relation to Higgs $\kappa$)
For a  type II dual superconductor
 $\kappa_d > 1/\sqrt{2}$ and a type I otherwise.
These simulations indicate that the non-Abelian dual superconductors
lie at the borderline between type I and type II.

Matsubara, Ejiri and
Suzuki\cite{mes} have measured $\kappa_d$ for SU(2) for $3 \times 3$
 and $5 \times 5$ loops in the range $2.4 < \beta  <2.6 $;
and for SU(3) for $3 \times 3 $ loop  in the range
$5.6 < \beta < 5.9 $.  Although they are not asymptotic, they find
clear signals in these data for $\kappa_d$ both larger and smaller
than $1/\sqrt{2}$.

The simulation\cite{ph} shown in fig.\ref{f5} was identical to fig.\ref{f4}
except that $\beta$ was increased to $2.40$
on a lattice $17^2 \times 19 \times 4$
putting us in the deconfining phase.
The dramatic decrease of $curl J^m$ results in the failure of
the dual superconductor interpretation as expected for the deconfining
phase.  Also $E$ falls more slowly with radius.

A further observation is that although this dual superconductivity
 signal drops
rapidly above the deconfining temperature, the monopole density
does not change dramatically. The monopole density $\rho$ is
defined
\begin{equation}
\rho = \frac{<\sum_{x ,\mu} |J^m_{\mu \;lat}|>}{4 N_{sites}}
\end{equation}
where $|J^m_{\mu \; lat}|$  is the integer valued lattice quantities and
$N_{sites}$ is the number of lattice sites.
In Ref.\cite{ph} we find
\begin{eqnarray}
confining: \hspace{.5in}  \beta &=&
2.28 \hspace{.5in} \rho = 0.0567(1) \nonumber \\
deconfining: \hspace{.5in}  \beta &=&
2.40 \hspace{.5in} \rho = 0.0213(1) \nonumber
\end{eqnarray}

\section{External Electric Field}
\label{efield}

Many interesting superconducting properties can be elucidated by studying
the properties of the material in the presence of a magnetic field. We
report a preliminary look at the corresponding problem of
dual superconductivity in the presence of an external electric field. The
Wilson loop provides such a field by projecting  a $q \bar{q}$
out of vacuum configurations.  But it may be interesting to see the
spontaneous breaking of translation invariance as the electric flux
forms a vortex rather than impose the vortex position at a given
location and with quantized flux.   Type I and type II dual
superconductors respond very differently to
background uniform external fields.

For a periodic U(1) lattice the sum of the plaquette angles over any
plane is identically zero.  Referring to Fig.\ref{ext}
a classical uniform electric field for U(1) on a particular time slice
 can still be obtained by
constraining one plaquette angle in each $z,t$ plane
to the value $(1/(N_zN_t)-1)\theta_c$.
Then the remaining $(N_z N_t-1)$
plaquette angles will take the common value
$\theta_c/(N_z N_t) = e a^2 F_{34}$ giving
a uniform electric field on all but one time slice. (N labels the lattice
dimensions.) Choosing $\theta_c = \pi$ gives the largest field.  For an
$8^4$ lattice and $\beta = 1$ for example, there would be enough electric
flux to form about 3 vortices in the $x,y$ plane in the dual superconducting
phase. This field configuration  can also be obtained by multiplying the
plaquette that was singled out in the action by a minus sign and such
configurations have been studied in non-Abelian
theories\cite{go}.  We avoided an
alternative method of imposing an external field by introducing a non-zero
equilibrium value locally for each plaquette angle since we eventually want
to see the field break translation invariance.
Turning on interactions for $\beta < 1$ brings up other interesting
features\cite{dh}.

Our goal here is to try to see a signal showing that $curl J^m$
responds to the external field.
The immediate problem is that the sum of $(curl J^m)_{x y}$
over the any $x,y$ plane is identically zero, and the sum of $E_z$ is not.
A sample configuration is given in Fig.\ref{zerosum}.
Yet
we expect the London relation to be satisfied.  The only possibility is that
translation invariance is broken which is expected since vortices segregate
the superconducting and normal phases.

We make the following rough ansatz that the local London relation is
due to a the alignment of local current loops in the external
field. This suggests that we truncate $curl J^m$ to include only
values $\pm 2,\pm 3,\pm 4$ representing the winding of the current around
the dual plaquette. We denote this $J^m_{\mu}(2+3+4)$. In this
discussion we assume
that the current $J^m_{\mu}$ takes only the values $-1,0,1$. The
fractional population
with values outside this range is small enough to be neglected.

Fig.\ref{zerosum} is a typical configuration for the confined phase.
A value $\pm 4$ indicates that the current makes a complete loop, the value
$\pm 3$ a partial loop, etc. The value $\pm 1$ corresponds to a single
isolated straight current segment in the plane.  One expects
the local behavior of $\pm 4, \pm 3, \pm2$ contributions to align
themselves in the local external field.  But there is no comparable local
behavior for the $\pm 1$ contributions.  Globally of course they must
all sum to zero.

On a large lattice where size effects are irrelevant, one can still
expect $curl J_m$ to average to zero.    This is precisely
the behavior in an isolated vortex as discussed in Sec.\ref{latticehiggs}.

Now we get a large signal for $curl J^m_{\mu}(2+3+4)$ as shown in
fig.\ref{f6} with a sign that agrees with the sign in fig.\ref{f3} for
the superconducting region, $r \neq 0$.  Further if we recalculate
fig.\ref{f3} with the truncated current, we find a large suppression at
$ r = 0$ and a moderate enhancement at the other points as shown
in Fig.\ref{f7}. In other words
this choice biases in favor of the dual superconducting phase.

\section{Acknowledgements}
I wish to thank D. Browne, A. Di Giacomo,
Y. Peng, M. Polikarpov, H. Rothe,
G. Schierholz, V. Singh, J. Wosiek, and K. Yee for many discussions.
I especially thank M. Polikarpov for many discussions on duality
transformations.  We
are supported in part by the US Department of Energy grant
DE-FG05-91ER40617.

\appendix{Notation Conventions}

\subsection{Lattice differential forms}
\label{diff}

The language of differential forms on the lattice is particularly
useful for the Villain\cite{villain}
form of the action. Fr\"{o}lich and
Marchetti\cite{fm} use this language in the mathematical
physics literature.  Polley and Wiese\cite[Sec. 2.1]{pw} have given a nice
introduction to the subject.  This appendix is meant only as a
supplement to that introduction
by relating the differential form operators to ordinary lattice
finite difference operators.   The differential form operators have
coordinate independent definitions\cite{pw}.
 The correspondences in this appendix allows
one to check the algebra of differential forms on a specific coordinate
system naturally defined by the hypercubic lattice.

The lattice forward and backward difference operators are defined
\begin{eqnarray}
  \Delta^{+}_{\mu}\Phi(x) \equiv \Phi(x + \epsilon^{(\mu)}) - \Phi(x), \\
  \Delta^{-}_{\mu}\Phi(x) \equiv \Phi(x) -\Phi(x - \epsilon^{(\mu)}).
\end{eqnarray}
We offer the following examples of the use of the
exterior differential d, and the codifferential $\delta$ acting on objects
defined on the original (rather than the dual) lattice.

0-form defined on points $\Phi(c_0) \Longleftrightarrow  \Phi(x)$:
\begin{eqnarray}
   d \Phi
  \hspace{.4in}  &\Longleftrightarrow&
 \hspace{.4in}   \Delta^{+}_{\mu}\Phi(x), \\
   \delta \Phi\hspace{.4in} &\Longleftrightarrow& \hspace{.4in}  0.
\end{eqnarray}

1-form defined on links $\Phi(c_{1}) \Longleftrightarrow \Phi_{\mu}(x)$:
\begin{eqnarray}
   d \Phi \hspace{.4in}  &\Longleftrightarrow& \hspace{.4in}
   \Delta^{+}_{\mu}\Phi_{\nu}(x) -
   \Delta^{+}_{\nu}\Phi_{\mu}(x), \\
   \delta \Phi \hspace{.4in} &\Longleftrightarrow& \hspace{.4in}
   \Delta^{-}_{\mu}\Phi_{\mu}(x).
\end{eqnarray}

2-form defined on plaquettes
$\Phi(c_{2})\Longleftrightarrow\Phi_{\mu \nu}(x)$ (antisymmetric):
\begin{eqnarray}
   d \Phi  \hspace{.4in}  &\Longleftrightarrow& \hspace{.4in}
   \Delta^{+}_{[\alpha}\Phi_{\mu\nu]}(x) \equiv\\
     && \hspace{.4in}
   \Delta^{+}_{\alpha}\Phi_{\mu\nu}(x)-
   \Delta^{+}_{\mu}\Phi_{\alpha\nu}(x)-
   \Delta^{+}_{\nu}\Phi_{\mu\alpha}(x),  \nonumber\\
   \delta \Phi \hspace{.4in} &\Longleftrightarrow& \hspace{.4in}
   \Delta^{-}_{\mu}\Phi_{\mu\nu}(x).
\end{eqnarray}

3-form defined on 3-volume $\Phi(c_{3})\Longleftrightarrow\Phi_{\mu \nu
\lambda}(x)$
(completely antisymmetric):
\begin{eqnarray}
   d \Phi \hspace{.4in}  &\Longleftrightarrow& \hspace{.4in}
   \Delta^{+}_{[\alpha}\Phi_{\mu\nu\lambda]}(x) \equiv\\
     && \hspace{.4in}
   \Delta^{+}_{\alpha}\Phi_{\mu\nu\lambda}(x)-
   \Delta^{+}_{\mu}\Phi_{\alpha\nu\lambda}(x)-
   \Delta^{+}_{\nu}\Phi_{\mu\alpha\lambda}(x)-
   \Delta^{+}_{\lambda}\Phi_{\mu\nu\alpha}(x),  \nonumber\\
   \delta \Phi \hspace{.4in} &\Longleftrightarrow& \hspace{.4in}
   \Delta^{-}_{\mu}\Phi_{\mu\nu\lambda}(x).
\end{eqnarray}

4-form defined on 4-volume $\Phi(c_{4})\Longleftrightarrow
\Phi_{\mu \nu \lambda \tau}(x)$
(completely antisymmetric):
\begin{eqnarray}
   d \Phi \hspace{.4in}  &\Longleftrightarrow& \hspace{.4in}   0,\\
   \delta \Phi \hspace{.4in} &\Longleftrightarrow& \hspace{.4in}
   \Delta^{-}_{\mu}\Phi_{\mu\nu\lambda\tau}(x).
\end{eqnarray}

Miscellaneous relations on k-forms:
\begin{eqnarray}
   \delta^2 = 0, \hspace{.4in}&& \\
    d^2 = 0, \hspace{.4in} &&\\
   \delta d + d \delta = \Delta  \hspace{.4in}
&\Longleftrightarrow& \hspace{.4in}
   \Delta^{-}_{\mu}\Delta^{+}_{\mu},\\
   \Phi = (\delta \Delta^{-1} d  + d \Delta^{-1} \delta)
   \Phi, \hspace{.4in}  &&  \\
   (\Phi,\Psi)\hspace{.4in}  &\Longleftrightarrow& \hspace{.4in}
   \sum_{x,\mu >\nu\cdots}
   \Phi_{\mu \nu\cdots}(x) \Psi_{\mu \nu \cdots}(x), \\
   (\Phi, d\Psi) =
   (\delta \Phi, \Psi). \hspace{.4in}  &&
\end{eqnarray}
Note that the two arguments of the inner product must of course live on
the same k-form.

There is a completely equivalent description of these k-forms,
 $\Phi(c_k)$,
in terms of (4-k)-forms on the dual lattice $^*\Phi(^*c_k)$.
\begin{equation}
   ^*\Phi(^*c_k) = \Phi(c_k).
\label{dddd}
\end{equation}
This is realized in conventional notation by contracting
$\epsilon_{\mu \nu \lambda \tau}$ into the object defined on the original
lattice to obtain the object on the dual lattice, properly normalized
so that acting twice is the identity operation.

Finally we wish to note a correspondence involving $d$ and $\delta$.
Given the equivalence Eqn.(\ref{dddd})
then there is a correspondence between the $(k \pm 1)$-form and
dual $(4-(k \pm 1))$-form
\begin{eqnarray}
d\Phi   =  \delta^*\Phi, \nonumber \\
\delta\Phi = d^*\Phi.
\end{eqnarray}

On the original lattice
the position $x$  refers to the `lower left etc.' corner of a k-form.
(Links point out of their coordinate address.)
 This dictates the choice of
$\Delta^+$ and $\Delta^-$ in these examples.  Since the dual lattice is
displaced a half spacing in the positive direction, the `upper right etc.'
corner is used to label the position $x$ of the dual k-forms.  (Dual links
point in to their coordinate address.)
As a consequence,  when the above
rules are applied to k-forms on the dual lattice, $\Delta^+$ and $\Delta^-$
should be interchanged.

\subsection{Euclidean Maxwell equations}
\label{maxwell}

We have taken the following sign conventions for the Euclidean
field tensors with electric and magnetic sources.
\begin{eqnarray}
\partial_{\mu} F_{\mu\nu}  &=& J^{e}_{\nu}, \;\;\;\;
^*F_{\alpha \beta} = -\frac{1}{2} \epsilon_{\alpha\beta\mu\nu} F_{\mu\nu},
\;\;\;\; \epsilon_{1234} = 1, \;\;\;\;
\partial_{\mu} ^*F_{\mu\nu} = J^{m}_{\nu}, \nonumber \\
\vec{\nabla} \cdot \vec{E} &=&
 J^e_4 , \;\;\;\; \vec{\nabla} \times \vec{B} -
\partial_4 \vec{E} = \vec{J^e}, \nonumber \\
\vec{\nabla} \cdot \vec{B} &=&  J^m_4,
 \;\;\;\; \vec{\nabla} \times \vec{E} -
\partial_4 \vec{B} = \vec{J^m}.
\end{eqnarray}
For $J^m_{\mu} = 0$ and with the standard definition $F_{\mu\nu} =
\partial_{\mu} A_{\nu}
- \partial_{\nu} A_{\mu}$, we get the unconventional sign $\vec{B} = -
\vec{\nabla} \times \vec{A}$.

\figure{DeGrand-Toussaint construction to identify a monopole in
a  spacial 3-volume.
\label{fdt}}

\figure{(a) Six space-time plaquettes (time not shown) contiguous
to a time-like link. (b) Six space-space plaquettes dual to those in
(a) forming a cube.
\label{fdual}}

\figure{Profile of $e a^2 F_{12}(x)$, Eqn.(\ref{e4}), and $
-(a^3/e \kappa)(\Delta^+_{1} J^{e}_{2}(x)
                    - \Delta^+_{2} J^{e}_{1}(x))$, Eqn.(\ref{e3}), as a
function of distance  perpendicular
to the plane of the monopole current loop in lattice units.
\label{f1}}

\figure{Operators for (a) $e a^2 F_{34}(x)$, and (b) $
(a^3/e_m)(\Delta^-_{1} J^{m}_{2}(x)
                    - \Delta^-_{2} J^{m}_{1}(x))$.
\label{curl}}

\figure{Profile of $e a^2 F_{34}(x)$ and $
-(a^3/e_{m})(\Delta^-_{1} J^{m}_{2}(x)
                    - \Delta^-_{2} J^{m}_{1}(x))$ as a function of
distance  perpendicular
to the plane of the Wilson loop in lattice units for U(1).
\label{f3}}

\figure{Profile of $e a^2 F_{34}(x) - \lambda_d^2
(a^3/e_{m})(\Delta^-_{1} J^{m}_{2}(x)
                    - \Delta^-_{2} J^{m}_{1}(x))$ as a function of
distance  perpendicular
to the plane of the Wilson loop projector in lattice units for U(1).
\label{f3a}}

\figure{Profile of $e a^2 F_{34}(x)$ and $
+(a^3/e_{m})(\Delta^-_{1} J^{m}_{2}(x)
                    - \Delta^-_{2} J^{m}_{1}(x))$ as a function of
distance  perpendicular
to the plane of the plaquette projector in lattice units. The on-axis
point for $curl J^m <0$ is `corrected', see text.
\label{f3b}}

\figure{Profile of $e a^2 F_{34}(x)$ and $
-(a^3/e_{m})(\Delta^-_{1} J^{m}_{2}(x)
                    - \Delta^-_{2} J^{m}_{1}(x))$ as a function of
distance  perpendicular
to the plane of the Polyakov lines separated by $R=3$
 in lattice units for SU(2) in the
confined phase.
\label{f4}}

\figure{Profile of $e a^2 F_{34}(x)$ and $
-(a^3/e_{m})(\Delta^-_{1} J^{m}_{2}(x)
                    - \Delta^-_{2} J^{m}_{1}(x))$ as a function of
distance  perpendicular
to the plane of the Polyakov lines separated by $R=3$
in lattice units for SU(2) in the
unconfined phase.
\label{f5}}

\figure{Plaquette angles for a classical external field configuration.
\label{ext}}

\figure{Sample values of the $z$ component of
$curl J^m_{lat}$ on an $x , y$ plane for
fixed $z$ and $t$ on a $12^4$ lattice, $\beta= 0.99$.
\label{zerosum}}

\figure{Average of $e a^2 F_{34}$ and $
-(a^3/e_{m})(\Delta_{1} J^{m}_{2}(2+3+4)
                    - \Delta_{2} J^{m}_{1}(2+3+4))$ on each time slice
for an external field $=\pi/64$ (the horizontal line).
The constrained plaquette is at $t=1$, the field is classical at
$t=2,8$ and $\beta=.99$ for $t=3-7$.
\label{f6}}

\figure{Profile of $
-(a^3/e_{m})(\Delta^-_{1} J^{m}_{2}(x)
                    - \Delta^-_{2} J^{m}_{1}(x))(2+3+4)$ as a function of
distance  perpendicular
to the plane of the Wilson loop in lattice units for U(1) superimposed
on Fig.\ref{f3}.
\label{f7}}

\end{document}